\documentclass[twocolumn,showpacs,floatfix,prl,superscriptaddress]{revtex4}

\usepackage{float}
\usepackage{graphicx}
\usepackage{bm}
\usepackage{amsmath}
\usepackage{amssymb}

\newcommand{\ta}{\tilde{\alpha}}

\newcommand{\beq}{\begin{eqnarray}}
\newcommand{\eeq}{\end{eqnarray}}
\newcommand{\beqq}{\begin{eqnarray*}}
\newcommand{\eeqq}{\end{eqnarray*}}

\begin{document}

\title{Entangled Cloning of Stabilizer Codes and Free Fermions}

\author{Timothy H. Hsieh}
\email{thsieh@kitp.ucsb.edu}
\affiliation{
Kavli Institute for Theoretical Physics, University of California, Santa Barbara, California 93106, USA}

\pacs{}

\begin{abstract}
Though the no-cloning theorem \cite{nocloning} prohibits exact replication of arbitrary quantum states, there are many instances in quantum information processing and entanglement measurement in which a weaker form of cloning may be useful.  Here, I provide a construction for generating an ``entangled clone'' for a particular but rather expansive and rich class of states.  Given a stabilizer code or free fermion Hamiltonian, this construction generates an exact entangled clone of the original ground state, in the sense that the entanglement between the original and the exact copy can be tuned to be arbitrarily small but finite, or large, and the relation between the original and the copy can also be modified to some extent.  For example, this work focuses on generating time-reversed copies of stabilizer codes and particle-hole transformed ground states of free fermion systems, although untransformed clones can also be generated.  The protocol leverages entanglement to simulate a transformed copy of the Hamiltonian without having to physically implement it and can potentially be realized in superconducting qubits or ultracold atomic systems. 

\end{abstract}

\maketitle

Entanglement both poses problems and offers solutions for quantum information processing.  On one hand, entanglement between a system and its environment leads to decoherence and makes the notion of a quantum memory challenging in practice.  On the other hand, entanglement features prominently in the solution of quantum error correction \cite{terhal, gottesman}, in which information is stored in logical bits which are entangled states of multiple physical qubits.   Such fault-tolerant stabilizer codes, in particular surface codes \cite{bravyi}, have progressed significantly in both theory \cite{dennis} and implementation \cite{fowler}.

Can entanglement be used to address other fundamental obstacles such as the no-cloning theorem, which forbids the replication of arbitrary quantum states?  Having multiple (approximate) copies of a quantum state would be useful for many purposes.  For example, schemes for measuring the $n$th Renyi entanglement entropy in quantum states, which have been recently proposed \cite{abanin, swap2} and realized \cite{islam}, require beginning with several ($n$) copies of the state.  Likewise, multiple copies of a state are directly useful for quantum state estimation and may serve other quantum information processing roles. 

The focus of this work are states which are the ground states of a Hamiltonian.  One trivial means of ``cloning'' is to replicate the Hamiltonian and thus the ground state, but in practice it may be challenging to duplicate the full Hamiltonian for the second copy, especially if the Hamiltonian is very complex.  To avoid doing so, I will make use of entanglement for this objective.  While most ``entanglement-assisted'' protocols for other purposes utilize maximally entangled states such as Bell pairs, I will utilize ``maximally entangling Hamiltonians'' which have a maximally entangled state as its ground state.  Such Hamiltonians are very common, as will be evident, and may be much simpler to implement than the full Hamiltonian whose ground state is to be replicated.       

In this work, I provide a construction for generating an ``entangled clone'' of any stabilizer code or free fermion Hamiltonian, without having to physically replicate the original Hamiltonian.  More precisely, given a stabilizer code, the output of this construction is the exact time-reversed copy of the original system, whose entanglement with the original system can be tuned to be arbitrarily small but finite, or large.  Similarly, given a free fermion Hamiltonian, the construction generates the exact particle-hole transformed copy of the original system, again tunably entangled with the original copy.  The construction can be modified so that time-reversal and particle-hole can be generalized to many other kinds of transformations, including no transformation (identical clones).  In the following, I detail the setup for this construction and state and justify the main claim of cloning (see equations (\ref{h1},\ref{h2}) and Fig. 1).  I conclude by discussing how this analysis applies to a wide range of systems and provides an entanglement perspective on the bulk topological proximity effect introduced in \cite{bpte}.       

\section{Setup and Definition}

Consider two identical Hilbert spaces: $A$, the parent system to be cloned, and $B$, an auxiliary system to realize the clone.  Let $H_A$ be a Hamiltonian for the $A$ system and let $H_{AB}$ describe coupling between $A$ and $B$ such that it has a unique ground state $|\psi\rangle$ which is maximally entangled between $A$ and $B$ (see Fig. 1 left).  
 
 The cloning results and proofs require the following definition:
 Given a state $|\psi\rangle$ which has maximal entanglement between two systems $A$ and $B$ and given an operator $O_A$ with support on $A$, the {\it dual operator} $O_B$ {\it relative to} $|\psi\rangle$ is defined as the operator which satisfies
\beq
O_A |\psi\rangle = O_B |\psi\rangle. \label{condition}
\eeq
Why is $O_B$ guaranteed to exist?  Due to maximal entanglement, $\rho_{A(B)} = tr_{B(A)} |\psi\rangle \langle \psi|$ is proportional to the identity matrix, which implies that
\beq
|\psi\rangle = \frac{1}{\sqrt{\cal N}}\sum_{\alpha} |\alpha\rangle |\ta\rangle,
\eeq 
where $\cal N$ is the size of the Hilbert space, $\{|\alpha\rangle\}$ is a complete, orthogonal basis for $A$, and $\{|\ta\rangle\}$ are corresponding states in $B$.  Then given $O_A$, $O_B$ is defined by the conditions
\beq
\langle \ta|O_B|\ta'\rangle = \langle \alpha' |O_A|\alpha\rangle \quad \forall \alpha, \alpha'.
\eeq
 One can easily check that these conditions ensure that (\ref{condition}) is satisfied.  Such dual operators relative to maximally entangled states are a special case of ``mirror operators" introduced \cite{mirror} in the black hole/holography context.  They allow one to re-express the action of an operator on one side of a maximally entanged state as the action of an operator acting on the other side.

\begin{figure}
\centering
\includegraphics[height=0.9in]{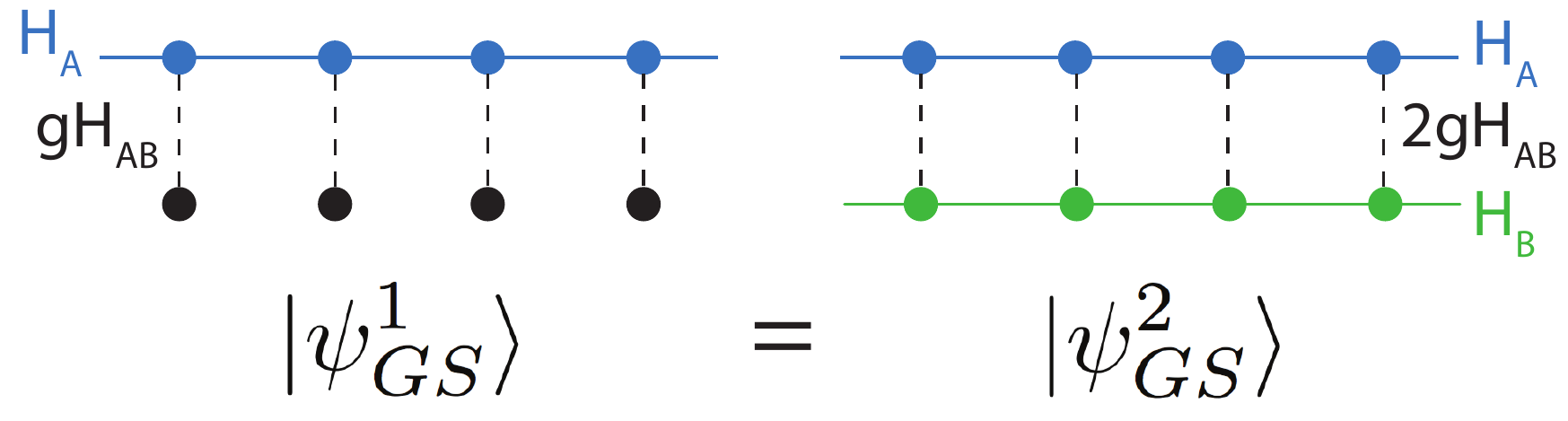}
\caption{(left) The Hamiltonian is $H_1 \equiv H_A + gH_{AB}$, where $H_A$ is a stabilizer code (blue system) and $H_{AB}$ is AFM exchange coupling between corresponding qubits of $A$ and $B$ (dashed lines). (right) The Hamiltonian is $H_1 \equiv H_A + H_B+2gH_{AB}$, in which $H_B$ is the time-reversed $H_A$ acting on the $B$ qubits and the coupling is twice as large.  The main claim is that both Hamiltonians have exactly the same ground state for all $g>0$.  This allows one to generate an entangled clone of the original ground state of $H_A$ without having to replicate the Hamiltonian.  A completely identical clone can be produced with ferromagnetic exchange (see (\ref{FM})).}
\end{figure}

\section{Cloning Stabilizer Codes}

In this section, let $A,B$ be two identical sets of $N$ qubits and $H_A$ be a stabilizer code:
\beq
&H_A = \sum_i H_{A,i} \\
&[H_{A,i},H_{A,j}]=0,
\eeq
where each operator $H_{A,i}$ is a string of Pauli operators $\sigma^A_{i,\alpha}$ acting on different sites $i$ ($\alpha=x,y,z$).  Notable examples include the cluster state \cite{cluster} and toric code \cite{toric} Hamiltonians, both of which figure prominently in quantum information processing proposals.  Moreover, I will consider the physically relevant exchange coupling 
\beq
H_{AB}=\sum_{i=1}^N \sigma^A_{i,x} \sigma^B_{i,x} + \sigma^A_{i,z} \sigma^B_{i,z} \label{exchange}
\eeq
Making the exchange isotropic by adding $\sigma^A_{i,y} \sigma^B_{i,y}$ will not alter the following conclusions.  Note that both the anisotropic and isotropic couplings have the same maximally entangled ground state $|\psi\rangle$, which is a product state of spin singlets formed from corresponding $A,B$ sites.  In this case, the dual operator of $H_A$, relative to $|\psi\rangle$ is 
\beq
H_B=H_A (\sigma^A\rightarrow -\sigma^B)
\eeq
This is because $\sigma^A_{i,\alpha} |\psi\rangle = -\sigma^B_{i,\alpha} |\psi\rangle$.

The main result is that the composite ground state $|\psi_1\rangle$ of 
\beq
H_1 \equiv H_A + gH_{AB}, \label{h1}
\eeq
for any coupling constant $g>0$, is exactly the same as the composite ground state $|\psi_2\rangle$ of 
\beq
H_2 \equiv H_A + H_B + 2gH_{AB}, \label{h2}
\eeq
This construction thus allows one to simulate $H_B$, a time-reversed $H_A$, without physically implementing it.

To establish the result, I start from the strong coupling limit, in which both Hamiltonians $H_1, H_2$ have the same ground state $|\psi\rangle$, and expand away from this limit to all orders.  By using the dual operator property of maximally entangled states, I will redistribute the action of $H_A$ onto both $A$ and $B$ and show that at every order, the ground states of $H_1$ and $H_2$ are identical.  The ground state of $\epsilon H_A + H_{AB}$ is given by 
\beq
|\psi_1\rangle=|\psi\rangle + \epsilon G_0 P_{\perp} H_A |\psi\rangle + \epsilon^2 (G_0 P_{\perp} H_A)^2 |\psi\rangle + O(\epsilon^3) \nonumber
\eeq
where $P_{\perp}\equiv 1-|\psi\rangle\langle \psi|$ and $G_0 \equiv (E_0-H_{AB})^{-1}$.

As a warmup, consider the $O(\epsilon)$ term.  Thanks to the dual operator property, which allows one to trade the action $H_A |\psi\rangle$ for $H_B |\psi\rangle$, we can equivalently write the term as
\beq
(\epsilon/2) G_0 P_{\perp} (H_A + H_B) |\psi\rangle.
\eeq

In fact, I will show that at any order $n$,
\beq
 (\epsilon G_0 P_{\perp} H_A)^n |\psi\rangle = \big((\epsilon/2) G_0 P_{\perp} (H_A + H_B)\big)^n |\psi\rangle \label{order}
\eeq

Expanding $H_A$ into its constituent operators transforms the left hand side into  
\beq
\epsilon^n \sum_{i_1,...i_n} \prod_{j=i_1}^{i_n} (G_0 P_{\perp} H_{A,j}) |\psi\rangle. \label{expansion}
\eeq

Moreover, by inspection one can check that $\sigma^A_{i,\alpha}|\psi\rangle$, for any $\alpha=x,y,z$, is also an eigenstate of $H_{AB}$.  Hence, in the above expression, every $G_0 P_{\perp}$ always acts on an eigenstate of $H_{AB}$ and thus reduces from an operator into a number, possibly zero.  As all $\{H_{A,j}\}$ commute with each other by assumption, each instance of $H_{A,j}$ can thus be commuted all the way to the right to act directly on $|\psi\rangle$, upon which it can be rewritten as $(H_{A,j}+H_{B,j})/2$, where $H_{B,j}$ is the dual operator of $H_{A,j}$ relative to $|\psi\rangle$.  The new operator $(H_{A,j}+H_{B,j})/2$ can then be commuted back to the original position of $H_{A,j}$.  This establishes the equivalence (\ref{order}) of the order $n$ terms in the expansions of $\epsilon H_A + H_{AB}$ and $(\epsilon/2)( H_A +H_B) + H_{AB}$.  

One can also expand from the weak coupling side.  In finding the ground state of $H_1 = H_A + gH_{AB}$, the degeneracy of $B$ is first lifted by an effective Hamiltonian $H_B$ that is the time reversed version of $H_A$; this is a consequence of (1) the integrable structure of $H_A$, (2) the fact that the $n$th order of degenerate perturbation theory carries a sign $(-1)^{n+1}$, and (3) $H_{AB}$ is antiferromagnetic (the coupling is positive).  Hence, at zeroth order in $g$, the ground state is identical to that of $H_2=H_A+H_B + 2gH_{AB}$.  At higher orders, the perturbative expansions again match exactly because any action of $H_{AB}$ flips the same integrals of motion in $A$ and $B$ (twice the energy cost is incurred in $H_2$ due to $H_B$, thus requiring twice the coupling to match with $H_1$).

All perturbative expansions converge for a parameter range as least $O(1/N)$ in size, given a gap of $H_A$ of $O(1)$.  However, given the equivalence at {\it all orders} in the expansion from {\it both} the weak and strong coupling limits, as well as support from numerical studies \cite{numerics} in the intermediate coupling regime, it is very suggestive that the ground states of $H_1$ and $H_2$ are equivalent for all $g$, not just the naive regime of convergence.  

\section{Cloning Free Fermions}

A similar result and proof holds for free fermion systems.  In this case, let $A$ and $B$ each be a set of $N$ fermions which can occupy half of $2N$ orbitals labeled $\sigma$.  Let $H_A$ be a noninteracting Hamiltonian which can be diagonalized into the form 

\beq
H_A = \sum_{\alpha} \epsilon_{\alpha} c^{\dagger}_{A,\alpha} c_{A,\alpha},
\eeq
where $\alpha (\epsilon_{\alpha})$ label eigenmodes(energies) of $H_A$.  In contrast to the exchange coupling chosen in the previous section, the coupling between $A$ and $B$ for fermions is chosen to be tunneling:
\beq
H_{AB} = -\sum_{\sigma} c^{\dagger}_{A,\sigma} c_{B,\sigma} + h.c.=-\sum_{\alpha} c^{\dagger}_{A,\alpha} c_{B,\alpha} + h.c., \nonumber
\eeq
which has a maximally entangled ground state: 
\beq
|\psi\rangle = \prod_{\alpha} (c^{\dagger}_{A,\alpha} + c^{\dagger}_{B,\alpha})|0\rangle
\eeq 
In this case, the dual operator of $H_A$, relative to $|\psi\rangle$, is 
\beq
H_B=H_A (c^{\dagger}_{A} c_{A} \rightarrow -c^{\dagger}_{B} c_{B})+ const.
\eeq
This is because $c^{\dagger}_{A,\alpha} c_{A,\alpha} |\psi\rangle = c_{B,\alpha}c^{\dagger}_{B,\alpha} |\psi\rangle=(1-c^{\dagger}_{B,\alpha} c_{B,\alpha}) |\psi\rangle$.

Once again, the assertion is that $H_1=H_A+gH_{AB}$ and $H_2=H_A+H_B+2gH_{AB}$ have identical ground states.  The argument is very similar to the one above, with a small difference: in this case, all intermediate eigenstates in the expansion (\ref{expansion}) involve either $(c^{\dagger}_{A,\alpha} + c^{\dagger}_{B,\alpha})|0\rangle$ or $(c^{\dagger}_{A,\alpha} - c^{\dagger}_{B,\alpha})|0\rangle$.  Both states are maximally entangled and, importantly, have the same dual operator correspondence.  Hence, every operator $H_{A,j}$ in the expansion directly acts on either of the intermediate eigenstates and can be transmuted in both cases into $(H_{A,j}+H_{B,j})/2$.  This establishes the claim.  

\section{Applications and Variants}

The two main results apply to stabilizer codes and free fermions, which encompass a wide range of states, including symmetry protected topological states (e.g. cluster states \cite{cluster}, topological insulators \cite{kanehasan}), topologically ordered states (e.g. toric code \cite{toric}, doubled semion model \cite{levinwen}), and exotic states like the Haah code \cite{haah1, haah2}.  All such states can be cloned in the entangled fashion above.

Moreover, fermionic stabilizer codes are also amenable to entangled cloning.  For example, recently studied models \cite{plaquette, haahmajorana} involve lattices of Majorana modes $\gamma^A_j$, whose Hamiltonian involves products of Majorana modes which mutually commute.  One can entangle clone such states by introducing an identical Hilbert space $B$ of Majorana modes $\gamma^B_j$ and coupling the subsystems with the Hamiltonian $H_{AB}=\sum_j i \gamma^A_j \gamma^B_j$.  This coupling has a ground state $|\psi\rangle$ which is maximally entangled with respect to the Majorana modes and operators can be dualized accordingly: $\gamma^A_j |\psi\rangle = i\gamma^B_j |\psi\rangle$.

The choice of couplings (exchange for spin, tunneling for fermions) dictates how the cloned system relates to the original system, and these aspects can be tailored in many different ways.  For example, while the exchange coupling considered gives rise to a time-reversed copy of $A$ for system $B$, an alternative coupling $\sigma^A_x \sigma^B_x - \sigma^A_z \sigma^B_z$ has the ground state $|\psi\rangle = |\sigma^A_z=1\rangle |\sigma^B_z=1\rangle - |\sigma^A_z=-1\rangle |\sigma^B_z=-1\rangle$ which admits a different duality $\sigma^A_x|\psi\rangle = -\sigma^B_x|\psi\rangle, \sigma^A_y|\psi\rangle = \sigma^B_y|\psi\rangle, \sigma^A_z|\psi\rangle = \sigma^B_z|\psi\rangle$.  In this sense, the coupling can be modified to produce different entangled clones.

In particular, the ferromagnetic coupling 
\beq
H_{AB}=-\sum_{i=1}^N \sigma^A_{i,x} \sigma^B_{i,x} + \sigma^A_{i,y} \sigma^B_{i,y} + \sigma^A_{i,z} \sigma^B_{i,z} \label{FM}
\eeq
can be used to produce untransformed clones ($H_B$ will be identical to $H_A$ in this case).   Similarly, in the free fermion case, pairing between $A$ and $B$
\beq
H_{AB} = \sum_{\sigma} c^{\dagger}_{A,\sigma} c^{\dagger}_{B,\sigma} + h.c.
\eeq
can be used to generate completely identical clones.

Moreover, depending on the coupling, the result may generalize beyond stabilizer codes.  For example, if $A$ and $B$ are two qubits coupled with isotropic Heisenberg exchange, then the main result applies for all single qubit Hamiltonians $H_A$ even though they may not be stabilizer codes.  This is because the rotational symmetry can be leveraged to rotate $H_A$ to $\sigma_z$, for which the stabilizer result applies.   

Finally, the arguments above can be readily generalized to justify equalities between the ground state of $H_1$ and the ground state of a continuous family of Hamiltonians: 
\beq
H_{\alpha} = (1-\alpha g)H_A + \alpha g H_B + gH_{AB}
\eeq
where $0<\alpha g < 1$ (the original case discussed is $\alpha g =1/2$).  Thus, there is an entire family of Hamiltonians with the same ground state of $H_1$ for arbitrary finite $g$.

\section{Bulk Topological Proximity Effect}

This cloning construction provides a new entanglement perspective of the bulk topological proximity effect introduced in \cite{bpte}, which I will now briefly review and revisit.  The setup considered in the prior work was also $H_1 = H_A + gH_{AB}$, where $H_A$ was assumed to be a topologically nontrivial system with gap $\Delta_A$ above the ground state, and the authors were primarily interested in the regime $g<<\Delta_A$.  It was established that, when $H_A$ is a free fermion system with a topologically nontrivial ground state and $H_{AB}$ is tunneling between corresponding degrees of freedom, the ``inverse'' topological phase is induced in system $B$ for arbitrarily small or large coupling $g$; more precisely, the entire system is topologically trivial even for arbitrarily small $g$.  Entangled cloning provides new insight into this phenomenon; it exactly maps the ground state of $H_1$ to the ground state of $H_2= H_A + H_B+2gH_{AB}$, upon which it is evident that system $B$ is already in the dual (in this case inverse) phase $H_B=H_A (c^{\dagger}_{A} c_{A} \rightarrow -c^{\dagger}_{B} c_{B})$ even for arbitrarily small $g$.  

The entangled cloning perspective is even more valuable for understanding the proximity effect of stabilizer codes. The prior work \cite{bpte} perturbatively analyzed the proximity effect of the toric code state for weak coupling $g$.  Unlike the free fermion topological phases, toric code hosts intrinsic topological order and, when weakly coupled to an identical auxiliary system, generates another copy of itself.  Instead of being trivial, the composite system is doubly nontrivial. This was concluded via perturbation theory from the weak coupling limit.  The entangled cloning provides a complementary analysis from the infinite coupling limit, concluding that the dual Hamiltonian $H_B$ is effectively simulated due to coupling alone.  In the case of toric code, all operators involve an even number of spins and thus the time-reversed Hamiltonian $H_B$ is identical to the original copy $H_A$. 

Finally, entangled cloning extends the proximity framework to gapless phases of $H_A$; unlike the original analysis there is no need to assume a gap $\Delta_A$. 

\section{Summary and Discussion}

I have provided a protocol which takes as an input a stabilizer code or free fermion Hamiltonian and outputs an exact entangled clone of the original ground state, whose entanglement with the original copy can be tuned.  In the specific examples illustrated above, the entangled clone is a time-reversed and particle-hole transformed copy of the original, but these particular transformations can be generalized using different maximally entangling couplings or avoided by using ferromagnetic/pairing coupling.  In addition to possible applications in quantum information processing/state tomography and measuring entanglement entropy, this construction provides a route to realizing new phases via coupling alone-- the simplest example being two copies of toric code from a single copy.  In particular, the realization of two toric codes coupled with ferromagnetic exchange may host interesting phases even in the strong coupling limit.  

Note that while free fermions and stabilizer codes are exactly solvable, adding the exchange coupling spoils the exact solvability of stabilizers.  Nonetheless, it is a pleasant surprise that the enlarged model still admits an exact duality at the ground state level to a related model, for a continuous range of couplings, thanks to the structure of maximally entangled states.  Moreover, the cloned stabilizer system constitutes a subsystem code (in which not all operators in the Hamiltonian commute and yet logical and stabilizer operators can still be defined), which may offer advantages in error correction \cite{bacon}.  While this work provides exact results for cloning stabilizer codes and free fermions, it would be useful to generalize to non-integrable Hamiltonians for system $A$, for which the exact mapping between ground states would likely relax to a mapping between phases; in other words, one still expects this construction to effectively clone the phase of $H_A$.    

Other interesting extensions from the ideal cases presented involve couplings which are entangled but not maximally entangled; these may still enable operator dualities but only for a subset of operators depending on the particular entangled state.  Such weaker couplings may admit cloning for a more restricted class of Hamiltonians.  Note however that maximal entanglement is sufficient but not necessary to have duality for all operators; for example, the thermofield double state \cite{tfd} $|\psi\rangle \propto \sum_{\alpha} e^{-\beta E_{\alpha}/2} |\alpha\rangle |\ta\rangle$ also allows dual operators $O_B$ to be defined via $\langle \ta|O_B|\ta'\rangle = e^{\beta (E_{\alpha'}-E_{\alpha})/2}\langle \alpha' |O_A|\alpha\rangle$.  See \cite{mirror} for many more details on such operator correspondences.

Superconducting qubits and ultracold atoms are two venues in which entangled clones of stabilizer codes and free fermions may be realized.  Hamiltonians involving both spins \cite{scspin} and fermions \cite{scfermion} have been successfully realized using superconducting qubits, and much potential remains (see e.g. \cite{potential1,potential2}).  Similarly, optical lattices feature highly tunable couplings and have shown significant progress  toward realizing topological phases of fermions \cite{review} and stabilizer codes \cite{ionqubit}; these are prime candidates for entangle cloning free fermion states, which may in turn facilitate the measurement of their entanglement entropies.  

\begin{acknowledgements}
{\it Acknowledgements:} TH thanks Leon Balents, Daniel Gottesman, Tarun Grover, G\'abor B. Hal\'asz, Wen-Wei Ho, Isaac Kim, Yuan-Ming Lu, Andreas Ludwig, Beni Yoshida, and Guanyu Zhu for interesting discussions.  TH is supported by a fellowship from the Gordon and Betty Moore Foundation EPiQS initiative (Grant 4304). 
\end{acknowledgements}


\begin{thebibliography}{10}
\bibitem{numerics}
Using an 11-site cluster state \cite{cluster} stabilizer Hamiltonian for $H_A$, the claim of equality of ground states of $H_1$ and $H_2$ has been checked for intermediate $g$.
\end{thebibliography}
\end{document}